\documentclass{aastex62}

\usepackage{amsmath}

\begin{document}

\title{The Virgo Overdensity Explained}

\author{Thomas Donlon II}
\affiliation{Department of Physics, Applied Physics and Astronomy, Rensselaer Polytechnic Institute, Troy, NY 12180, USA}
\author{Heidi Jo Newberg}
\affiliation{Department of Physics, Applied Physics and Astronomy, Rensselaer Polytechnic Institute, Troy, NY 12180, USA}
\author{Jake Weiss}
\affiliation{Department of Physics, Applied Physics and Astronomy, Rensselaer Polytechnic Institute, Troy, NY 12180, USA}
\author{Paul Amy}
\affiliation{Department of Physics, Applied Physics and Astronomy, Rensselaer Polytechnic Institute, Troy, NY 12180, USA}
\author{Jeffery Thompson}
\affiliation{Department of Physics, Applied Physics and Astronomy, Rensselaer Polytechnic Institute, Troy, NY 12180, USA}
\affil{Division of Natural Sciences and Mathematics, Southern Vermont College, Bennington, VT 05201, USA}

\begin{abstract}
We suggest that the Virgo Overdensity (VOD) of stars in the stellar halo is the result of a radial dwarf galaxy merger that we call the Virgo Radial Merger.  Because the dwarf galaxy passed very near to the Galactic center, the debris has a large range of energies but nearly zero $L_z$ angular momentum.  The debris appears to extend from 5 to 50 kpc from the Sun in the Virgo region.  We connect different parts of this merger debris to the Perpendicular and Parallel Streams (the Virgo Stellar Stream is associated with either or both of these streams), the Hercules-Aquila Cloud (HAC), and possibly the Eridanus Phoenix Overdensity (EriPhe). This radial merger can explain the majority of the observed moving groups of RR Lyrae and blue horizontal branch stars that have previously been identified in Virgo. This merger also produces debris in the Solar neighborhood similar to that identified as the {\it Gaia}-Enceladus or {\it Gaia}-sausage merger.  Orbits are provided for components of the Virgo Radial Merger progenitor and for debris that appears to be related to the Cocytos Stream, which was also recovered in the Virgo region. 
\end{abstract}


\section{Introduction} \label{sec:intro}

The Virgo Overdensity (VOD) was originally identified by \cite{Vivas2001} as a clump, called the ``12$^{\rm h}$.4 clump,'' of RR Lyrae stars from the Quasar Equatorial Survey Team (QUEST) survey.  It was also evident as a significant halo substructure, dubbed S297+63-20., in the equatorial slice of Sloan Digital Sky Survey \citep[SDSS;][]{SDSSYork} main sequence stars analyzed by \cite{Newberg2002}. 

\cite{Majewski2003} also identified an excess of 2MASS M giant stars in the Virgo constellation, calling it a ``descending, foreshortened northern loop," suggesting that it is part of the leading tidal tail of the Sagittarius (Sgr) dwarf galaxy.  The possible connection was amplified by \cite{MartinezDelgado2007}, who pointed out that the \cite{Law2005} model for the Sgr dwarf tidal tails passed through the region of the Virgo Overdensity; the suggestion was that the leading tail of the Sgr dwarf spheroidal galaxy was coming straight down towards the Galactic plane right at the position of the Sun. 

In a density study of the north Galactic cap in SDSS (primarily main sequence) stars using photometric parallax, \cite{Juric2008} identified a halo structure covering more than 1000 sq. deg. of sky, at distances of 6-20 kpc from the Sun, and named it the Virgo Overdensity.  Because this paper was submitted in 2005 and made publicly available at that time, the name and results are referred to in papers that appeared in the published literature much earlier than its publication date.

\cite{Duffau2006} identified both RR Lyrae and BHB stars in the structure, noting that 6 of the 9 stars in the densest region had line-of-sight, Galactic standard-of-rest velocities of $V_{GSR}\sim 100$ km s$^{-1}$; this was thought to be debris from a tidally disruped, low-luminosity dwarf spheroidal galaxy and the name ``Virgo Stellar Stream (VSS)" was suggested.  The VSS is located $\sim$20 kpc from the Sun and covers more than 100 square degrees of the sky.  Curiously, the histogram of line-of-sight velocities in Figure 2 of this paper showed peaks at both $V_{GSR} \sim 100$ km s$^{-1}$ and $V_{GSR} \sim -100$ km s$^{-1}$, perhaps suggesting the existence of more substructure in the VOD region than just the VSS.

\cite{Newberg2007} measured the line-of-sight velocity of the S297+63-20. structure (which was corrected to S297+63-20.5 in this paper) as $V_{GSR}=130$ km s$^{-1}$.  It appeared likely that this SDSS structure was the same as the VSS (since it had a similar position and a velocity that differed by only 30 km s$^{-1}$), and to the VOD (though the VOD seemed to possibly be extended and somewhat closer to the Sun).  The positive line-of-sight velocity (indicating the stream was moving away from us rather than falling into the disk from above) and a difference in turnoff color between S297+63-20.5 and the Sgr dwarf tidal stream ruled out a connection between these two structures.  The Sgr dwarf tidal stream was also shown to miss the position of the Sun by 15 kpc when it passed through the disk.   It is interesting to note that in this paper as well \citep[see lower left panel of Figure 10 in][]{Newberg2007} there is also evidence of both positive and negative line-of-sight, Galactic standard-of-rest velocities $-$ in this case for stars that were thought to be slightly bluer and closer than the VSS.

In the following years the VOD/VSS structure continued to be explored, but no major progress was made in understanding its nature and the relationship between the VOD and VSS.  Many papers found moving groups \citep{Vivas2008, Prior2009, Brink2010, Casey2012} and even density substructures \citep{Keller2009}.  The photometrically determined metallicity was explored in \cite{An2009}.  Authors continued to find both negative and positive line-of-sight velocities in the VSS/VOD region; typically debris with a distance of ~20 kpc and a $V_{GSR}\sim 130$ km s$^{-1}$ was called the VSS, and debris that ranged over a larger distance range (10-20 kpc) was called the VOD.  Authors tried to associate the stars with negative $V_{GSR}$ with the leading tidal tail of Sgr, but no convincing connection was made.


\cite{CasettiDinescu2009} measured the proper motion of a single RR Lyrae star in the VSS, and from that derived an orbit that indicated that this substructure was the result of a dwarf galaxy accretion event that was found near the pericenter of a highly eccentric (and destructive) orbit.  This result was amplified by \cite{Carlin2012}, who found a similar orbit using 16 stars with measured proper motions that were plausibly part of the VSS.  Evidence was building that the VSS/VOD covered 2000 or even 3000 sq. deg. of sky, and could be a single merger event on a highly eccentric orbit, discovered near perigalacticon \citep{Bonaca2012}.  \cite{Jerjen2013} noted that the VSS/VOD have similar main sequence stars, but that the VSS/VOD at 23 kpc was nowhere near the Sgr stream and was therefore not associated.  They measured an age of 8.2 Gyr and an abundance of -0.67 dex for these stars.


However, the single progenitor idea did not last.  \cite{Duffau2014, Vivas2016} found several halo substructures along the same line of sight in Virgo.  \cite{Zinn2014} noted that the VSS was at the most distant part of the much larger VOD structure, but pointed to the large number of identified moving groups as evidence that the structure might not arise from a common origin.


\citet{Weiss2018a,Weiss2018b} attempted to characterize the density substructure of the north Galactic cap, including the Sgr tidal tails and the VOD, using statistical photometric parallax and a large sample of SDSS turnoff stars. However, they discovered that there was more substructure in this region than expected.  In particular, they identified a density substructure that might be associated with a moving group called the Parallel Stream that was discovered by \cite{Sohn2016}. They also identified a new stream oriented vertically with an RA $\sim 190^\circ$ that was named the Perpendicular Stream.  These two streams cross in the densest region of the VOD, and it was suggested that the VSS/VOD might be primarily composed of these two substructures.  The Perpendicular Stream was independently discovered in RR Lyrae stars by \cite{Boubert2019}, who suggested this structure could be associated with the Magellanic Clouds or the Vast Polar Structure.


The availablility of data from {\it Gaia} Data Release 2 \citep[DR2;][]{GaiaCollaboration2018} has ushered in a new era in understanding the substructure of the Milky Way stellar halo and disk.  In particular, a significant halo substructure has been identified that is thought to be the remains of a dwarf galaxy merger that came into the Milky Way on a highly radial orbit 10 Gyr ago \citep{Helmi2018, Belokurov2018}.  This structure has been variously called the {\it Gaia}-Enceladus merger or the {\it Gaia}-sausage, and it has been proposed that the Hercules-Aquila Cloud (HAC) in the south and the VOD could be part of that same ancient accretion event \citep{Simion2019, Iorio2019}.  The Eridanus-Phoenix (EriPhe) overdensity is also possibly connected to the VOD and HAC \citep{Li2016}.

In this paper we match stars in the moving groups discovered by \cite{Duffau2014} and \cite{Vivas2016} to {\it Gaia} proper motions, and show that these groups and the VOD could be largely explained by a single merger event on a highly radial orbit (which we call the Virgo Radial Merger), plus a stream whose properties are consistent with the Cocytos Stream \citep{Grillmair2009}.  Since the highly radial merger produces debris at a range of energies, different parts of the merger event can be associated with the Perpendicular and Parallel Streams.  The radial merger can also account for both the positive and negative radial velocities observed in the VSS/VOD; since the stars are on radial orbits; one sees the stars going towards apogalacticon and returning from apogalacticon in the same direction in the sky. It is reasonable that the merger has been previously identified as many small groups in distance and line of sight velocity, because a radial merger produces a wide spread of negative and positive line of sight velocities not typically seen in known stellar streams.  The Parallel Stream is associated with the VSS orbit as fit by \cite{Carlin2012}, but other identifications of VSS structure could be associated with either or both of the Parallel/Perpendicular Streams. We provide orbits for substructure with three different energies, found at distances of 5-20 kpc in Virgo.

Our exploration suggests that the {\it Gaia}-Enceladus/{\it Gaia}-sausage structure \citep{Helmi2018, Belokurov2018, Lancaster2018} could be related to the Virgo Radial Merger.  Both substructures are highly radial and occupy a similar regions of Galactocentric rotational velocity vs. Galactocentric radial velocity. However, we are able to create a substructure that qualitatively looks like the VOD with a 2 Gyr simulation.  This is far more recent than has been proposed for the {\it Gaia}-Enceladus/{\it Gaia}-sausage merger.  We do not fully explore the age of the Virgo Radial Merger or whether the halo is composed of one or more radial mergers. However, if we remove the RR Lyraes associated with the Virgo Radial Merger from our sample, we do not see a remaining radial merger remant in the portion of the halo that is 5-20 kpc from the Sun in the direction of Virgo.

\section{Moving group catalogs} \label{sec:data}

In this work, we look at two catalogs of halo moving groups from the region of the VSS/VOD.  These catalogs contain RR Lyrae and blue horizontal branch (BHB) stars from \cite{Duffau2014} and RR Lyrae stars from \cite{Vivas2016}; they are publicly available on VizieR \citep{Ochsenbein2000}.  The catalogs provide  line-of-sight velocities and 3D positions for the stars identified with moving groups in these two papers.  Since these catalogs were released before {\it Gaia} DR2, they do not include proper motions.  We therefore cross-matched the moving group catalogs with {\it Gaia} DR2 using the TOPCAT software \citep{TOPCAT} to get the full 6D phase space information for all of the stars.  The cross-match was done in right ascension (RA) and declination (Dec) using a match radius of one arcsecond; a match was found for 89\% of the stars between both surveys.  The distances to the RR Lyrae stars were provided in the moving group catalogs. Using these distances, the proper motion data was then corrected for the solar reflex motion using the method described in \cite{JohnsonSoderblom1987}.

The \citet{Duffau2014} catalog consists of stars from five moving groups, named Group A, B, D, F, and H.  The stars in each group share similar heliocentric distances and line-of-sight velocities. \cite{Duffau2014} suggested that Group A was consistent with the Virgo Stellar Stream (VSS), but did not elaborate on the origins of the other groups. Out of the 5 groups discussed in that work, Group A contains the most stars, followed closely by Group B.  The \cite{Vivas2016} catalog contains 22 groups, also consisting of stars with similar heliocentric distances and line-of-sight velocities.

\section{The moving groups in 6D} \label{sec:6D}
\begin{figure}
\includegraphics[width=\linewidth]{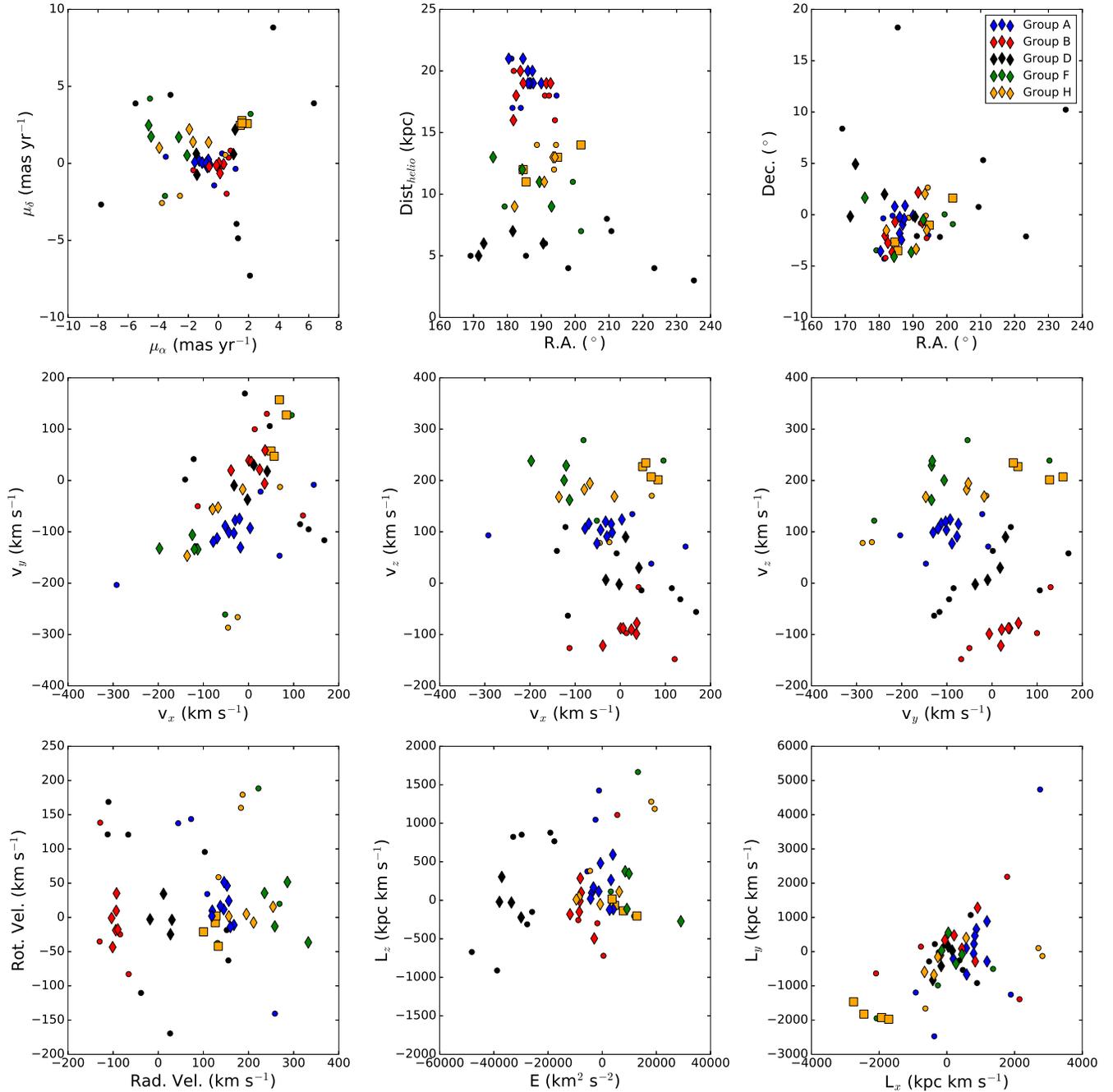}
\caption{Six dimensional phase space information and integrals of motion space information for all stars in five groups from the \cite{Duffau2014} catalog. Stars are colored by the group to which they belong (Group A, blue; Group B, red; Group D, black; Group F, green; and Group H, yellow).  The diamonds indicate the stars from each group that are comoving in $V_x, V_y, V_z$.  In Group H two moving groups were identified; we have associated the yellow squares with the Cocytos Stream, and the yellow diamonds with the Perpendicular Stream. Note that all of the co-moving stars have low rotational velocity and $L_z$ angular momentum.  While the different groups cover a range of energies, only all of the stars except those associated with Cocytos have similar $L_x, L_y$ angular momentum. \label{fig:starplot}}
\end{figure}

Using 6D phase space information, we reevaluated each of the five \cite{Duffau2014} moving groups to determine whether they were still in fact moving groups when proper motions are included. Stars were determined to be in a moving group if their proper motions and $V_x$, $V_y$, $V_z$ velocities appeared to be comoving with the rest of the group.  The top left panel of Figure \ref{fig:starplot} shows the proper motions of the RR Lyrae stars, color coded by moving group.  There is an obvious concentration of stars for Group A (blue) and Group B (red); the stars that are clustered in proper motion for these two groups are shown as filled diamonds in this panel.  The top center panel of Figure \ref{fig:starplot} shows that the RR Lyraes represented by the blue and red diamonds are about 20 kpc from the Sun and the top right panel shows that they are at approximately the same position in the sky.  The second row of panels in Figure \ref{fig:starplot} shows that the velocities of these two moving groups are very different, explaining why they were not grouped together in the original catalog.  \cite{Duffau2014} lists the line-of-sight, Galactic standard-of-rest velocity of Group A as 132 km s$^{-1}$, and Group B as -93 km s$^{-1}$.  Note that these are exactly the two line-of-sight velocities, for stars at exactly the same distance, found in \cite{Duffau2006}; however, only the positive velocity stars were attributed to the Virgo Stellar Stream at that time, while the Virgo Overdensity was still considered to be a ``cloud'' of many different moving groups with unknown origins.  

In the lowest row of Figure \ref{fig:starplot} we show the Galactocentric rotational velocity vs. radial velocity, the $L_z$ angular momentum vs. energy, and the $L_y$ angular momentum vs. the $L_x$ angular momentum. In this work we utilize a right-handed, Galactocentric Cartesian coordinate frame, where the $X$-axis points from the Sun to the Galactic center, the $Y$-axis points in the direction of the Sun's rotation around the Galaxy, and the $Z$-axis is the cross product of the $X$ and $Y$ axes pointing towards the north Galactic pole.  The energy was calculated using a logarithmic halo, Miyamoto-Nagai disk, and Hernquist bulge with parameters found in model 5 of Table 3 from \cite{Newberg2010}.  This potential goes to zero at about 34 kpc from the Galactic center in the $Z=0$ plane; positive energies in Figure \ref{fig:starplot} indicate an apogalaction of greater than 34 kpc and not necessarily that the star is unbound from the Galaxy.  The lowest row of Figure \ref{fig:starplot} shows that both the red and blue groups have zero rotational velocity, and thus zero $L_z$ angular momentum.  Groups A and B also have fairly similar energies and $L_x$, $L_y$ angular momenta.  We therefore suggest that these two most significant groups in the \cite{Duffau2014} catalog are part of the same stream, which is nearly radial.  The Group A stars are on their way out to apogalacticon and the Group B stars are coming back towards us from apogalacticon.

Group D is very close to us ($<$ 8 kpc away), with an average line of sight velocity of 30 km s$^{-1}$.  Because this group is close, it does not cluster much in proper motion.  However, we can identify four stars in this group that have similar $V_x, V_y, V_z$, and are depicted with filled black diamonds in Figure \ref{fig:starplot}.  Like Groups A \& B, this group is on a radial orbit with essentially zero rotational velocity and $L_z$.  In the bottom center panel of Figure \ref{fig:starplot}, it is shown that this group has a much lower energy than the other moving groups. In the lower right hand panel, it is shown that these stars have similar $L_x$ and $L_y$ angular momenta to those in Groups A \& B.

Group F stars are about 13 kpc from the Sun.  They only loosely clump in proper motion.  Four of the stars have been identified that have similar proper motions and $V_x, V_y, V_z$ velocities, and are depicted with filled green diamonds in Figure \ref{fig:starplot}.  These stars also have essentially zero rotational velocity and $L_z$ angular momentum.  We noted that there are also four stars in Group H that have similar velocities; these Group H stars are shown as filled yellow diamonds in Figure \ref{fig:starplot}.  The stars shown as green and yellow filled diamonds have similar angular mometum to those shown with red and blue filled diamonds, though there are small offsets between each group.  The Group F stars have a line of sight velocity of 215 km s$^{-1}$, while the selected Group H stars have a line-of-sight velocity of 171 km s$^{-1}$ and also share a heliocentric distance between 8 and 13 kpc with Group F.  Because the group H stars actually have energies more similar to the Group A \& B stars, we switched their association from F to A \& B even though the distance and velocities initially looked more similar to F.  Group F stars have the highest energy of all of the moving groups, suggesting that these stars are the least bound. 

Group H (yellow in Figure \ref{fig:starplot}), in addition to the four stars that look similar to Group F but are associated with Groups A \& B in energy, has four stars (yellow squares) with extremely similar proper motions, fairly similar $V_x, V_y, V_z$, and an $L_x$ below -1000 kpc km $s^{-1}$. These stars are also on a highly radial orbit (essentially zero rotational velocity and $L_z$), but have much different $L_x, L_y$ angular momentum than the other moving groups. The energy of this moving group appears to overlap with Group F.

In summary, we were able to identify co-moving sets of stars in all five of the RR Lyrae groups identified by \citet{Duffau2014}.  About half of the stars in each group (a larger fraction in Group A) were found to be comoving.  The comoving stars in all five groups are on radial orbits with essentially zero rotational velocity and $L_z$ angular momentum.  These comoving stars span a range of energies, but,  with the exception of four stars in Group H in a tight moving group, they have roughly similar $L_x, L_y$ angular momentum.  There appear to be somewhat more stars with rotational velocities of 150 km s$^{-1}$ than expected in the lower left panel of Figure \ref{fig:starplot}.  These stars are moving in the opposite direction from disk stars.  With the moving group stars (filled diamonds and squares) removed, there is no residual concentration of stars on radial orbits.



\section{Moving group orbits and N-bodies} \label{sec:orbits}

\subsection{The Perpendicular Stream} \label{sec:perpstream}

In the bottom center panel of Figure \ref{fig:starplot}, Group A and Group B (red and blue, respectively) overlap in integrals of motion space, suggesting that they may be related structures. We suspect that Group A and Group B are both members of a stellar stream traveling radially along our line of sight; Group A is the portion of the stream that is traveling away from us, and Group B is the portion of the stream that is traveling towards us. These two moving groups are found at the same position and distance because the stream is turning around in its orbit in this region of the sky, causing it to move through the same region twice with opposite radial velocities. 

A single orbit was fit to Groups A \& B together.  Because we expect that the orbit to turn around in heliocentric distance at a particular RA and Dec, the line-of-sight velocity would be double-valued in that direction. Typical orbit-fitting gradient descent algorithms \citep[such as the one described in][]{Martin2018}, which fit stream properties as a function of heliocentric angle in the sky to a calculated orbit, are not useful for fitting orbits that are directed towards or away from us along our line of sight, because they cannot determine which velocity to fit to each orbit point.  In order to fit stream properties that are single valued as a function of position along the stream, we fit the orbit in a Galactocentric frame.

In a Galactocentric frame, orbits lie roughly on planes that pass through the Galactic center (this is exactly true in a spherically symmetric Galactic potential). We begin by fitting a plane to the observed data using an ordinary least mean squares approach adapted from \cite{Newby2013}, while also making sure that this plane passes through the center of the Galaxy. We then convert our observed data into Galactocentric stream coordinates ($\Lambda$, $B$). $\Lambda$ is defined as the stream longitude coordinate (measuring angle along the stream) and $B$ is defined as the stream latitude coordinate (measuring angle from the orbital plane, orthogonal to $\Lambda$). Within this new frame, orbit properties are not double valued for a particular $\Lambda$, allowing us to fit the distance, sky position, and velocity data as a function of $\Lambda$. $\Lambda$ increases in the same direction as the Galactic longitude $l$, and $\Lambda$ = 0$^\circ$ at the Galactocentric (right-handed) X-axis projection in the $(\Lambda,B)$ plane.  

We generated orbits using Galpy version 1.3.0 \citep[][http://github.com/jobovy/galpy,]{Bovy2015} with a gravitational potential consisting of a Hernquist bulge, Miyamoto-Nagai disk, and a logarithmic halo using potential parameters from Orphan Stream Model 5 in \cite{Newberg2010}.  We then converted the data and the points along the orbit to $(\Lambda, B)$.

A $\chi^2$ function similar to the one presented in \cite{Willett2009} was used to determine the quality of fit of the orbit to the data.  This function compares the observed data (3D velocity and 3D position) for each of the stream stars with the point on the calculated orbit that is closest to that star's $\Lambda$ value. A small difference between the calculated orbit and the data corresponds to a small $\chi^2$ value, so the $\chi^2$ function was minimized in order to optimize the orbit. The potential used to evolve the orbit is the same one used to calculate the energies in Section \ref{sec:6D}. The code for this procedure can be found at \url{https://github.com/thomasdonlon/gc_orbit_fitting}.

\begin{figure}
\center
\includegraphics[width=0.5\linewidth]{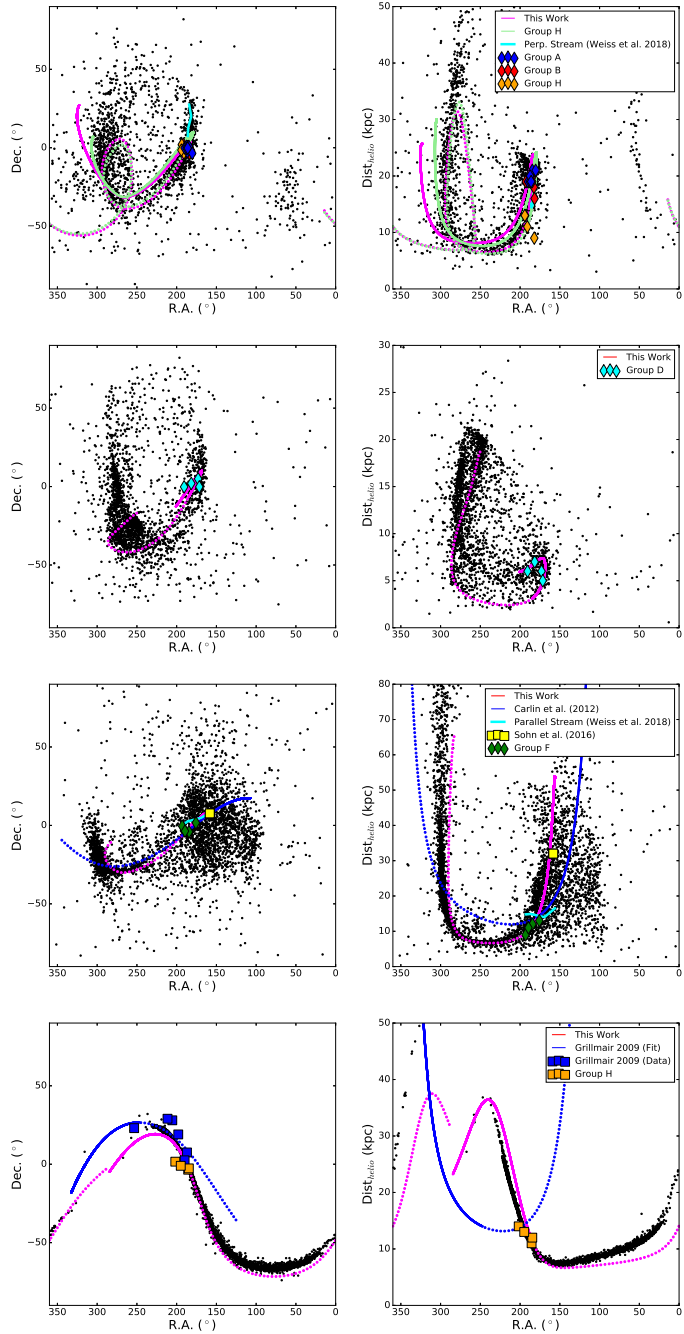}
\caption{ N-body simulations for each of the moving groups from \cite{Duffau2014}. Our calculated orbits are plotted in magenta on top of the N-bodies. All simulations were evolved for 2 Gyr, a progenitor mass of 10$^7 M_\odot$, a scale radius of 0.4 kpc, and 10,000 bodies. A subsample of 25\% of the total particles in the simulation are plotted in this figure. In the top two panels we plot the result of the N-body simulation for Groups A \& B (magenta).  We also show the orbit of a portion of the stars in group H (light green).  We also show the position of the Perpendicular Stream \citep[cyan;][]{Weiss2018b}.  Note that the two orbits are extremely similar, suggesting all of the stars are part of the same stream, and that they line up in the sky with the Perpendicular Stream.  In the second row of panels we show the orbit and N-body simulation for Group D.  The orbit and position of the debris is similar to the top two panels, but the debris does not go out as far due to the lower energy of these stars. The third row of panels shows the result of the N-body simulation for Group F (VSS/Parallel Stream).  Also shown is the orbit for the VSS (blue) from \citet{Carlin2012} and the fit to the Parallel Stream \citep[cyan;][]{Weiss2018b}.  Note that these are plausibly the same structure. Finally, in the bottom two panels, we plot an N-body simulation of Group H. For reference we also provide the \cite{Grillmair2009} polynomial fit for Cocytos (blue), which is similar in position on the sky, and at the same distance in the portion of the orbit over which the stream was observed.\label{fig:nbody}}
\end{figure}

The final optimized orbit values for the fit to Groups A \& B can be found in Table \ref{tab:orbit-params}.  The orbit fit to these two groups is shown in magenta in the top two panels of Figure \ref{fig:nbody}.  Also shown in the top panels, in light green, is the orbit defined by the average position and velocity of the stars in Group H that appear to be similar to those in Groups A \& B.  The similarity of the orbits supports the idea that these Group H stars are also from the same structure.  This orbit is not reported in Table \ref{tab:orbit-params} due to its similarity to the orbit for Groups A \& B.

We then ran an N-body simulation in the Galactic potential from Section \ref{sec:6D}, using the orbital parameters fit to Groups A \& B in Table  \ref{tab:orbit-params}. We used the MilkyWay@home N-body simulator to run our simulations \citep{SheltonThesis}. The dwarf galaxy progenitor for each simulation was modeled by a 10$^{7} M_{\odot}$ Plummer sphere density profile composed of 10,000 bodies of equal mass and a scale radius of 0.4 kpc. Each simulation was run by placing a single body at a position within the VOD at the location specified for each structure in Table  \ref{tab:orbit-params}.  This body was integrated backwards along the orbit for 2 Gyr.  The 10,000 bodies of the simulated dwarf galaxy progenitor are placed at the new position of the body and then integrated forwards for 2 Gyr. This procedure is guaranteed to place debris where we observe it in the VOD.  The N-body parameters provide a rough idea for the expected shape of the tidal stream, but there was no particular reason to select a progenitor of these properties.  

Note that the N-body places many particles at a RA of $180^\circ$ in the northern hemisphere, along a wide range of distances.  This is exactly the location of the Perpendicular Stream, shown as a cyan line \citep{Weiss2018b} in the upper panels of Figure \ref{fig:nbody}.  The shape of the stream is very similar to the shape of RR Lyrae stars in the northern sky depicted in \cite{Boubert2019}, an independent discovery of this stream.  We therefore suggest that Groups A, B, and part of H are associated with the Perpendicular Stream.  

Based on the large portion of the moving group data that belongs to the Perpendicular Stream \citep[Groups A and B are the richest groups in the ][catalog]{Duffau2014}, we assume this stream is a major contributor to total mass within the VOD. Since this stellar stream is approaching apogalacticon within the VOD region, it clumps up and provides a large amount of material in this region.  Note also that there are many N-body simulated particles in the southern hemisphere at approximately the location of the HAC (centered near RA, Dec = 285$^\circ$, +6$^\circ$), at a range of distances between 10 and 40 kpc.  

\subsection{The Horns of Virgo} \label{sec:horns}


Previous studies have observed the ``Horns of Virgo," a double-peaked distribution in $V_{GSR}$ within Virgo \citep{Duffau2006, Newberg2007, Duffau2014}.  Histograms of the $V_{GSR}$ of the stars in the VOD typically show peaks near $V_{GSR}=-75$ km s$^{-1}$ and $V_{GSR}=130$ km s$^{-1}$, with slight variations depending on the region of the VOD being observed and the distance to the stars in the study.  Based on these peaks, it was assumed that there were at least two different substructures within the VOD.  Instead, we show that the two peaks belong to the same substructure that is receding to aphelion and then approaching after having passed through aphelion, with both the outgoing and incoming stars moving in directions similar to our line-of-sight.  By this logic, there are more distant stars that have zero $V_{GSR}$ velocity.

\begin{figure}
\center
\includegraphics[width=\linewidth]{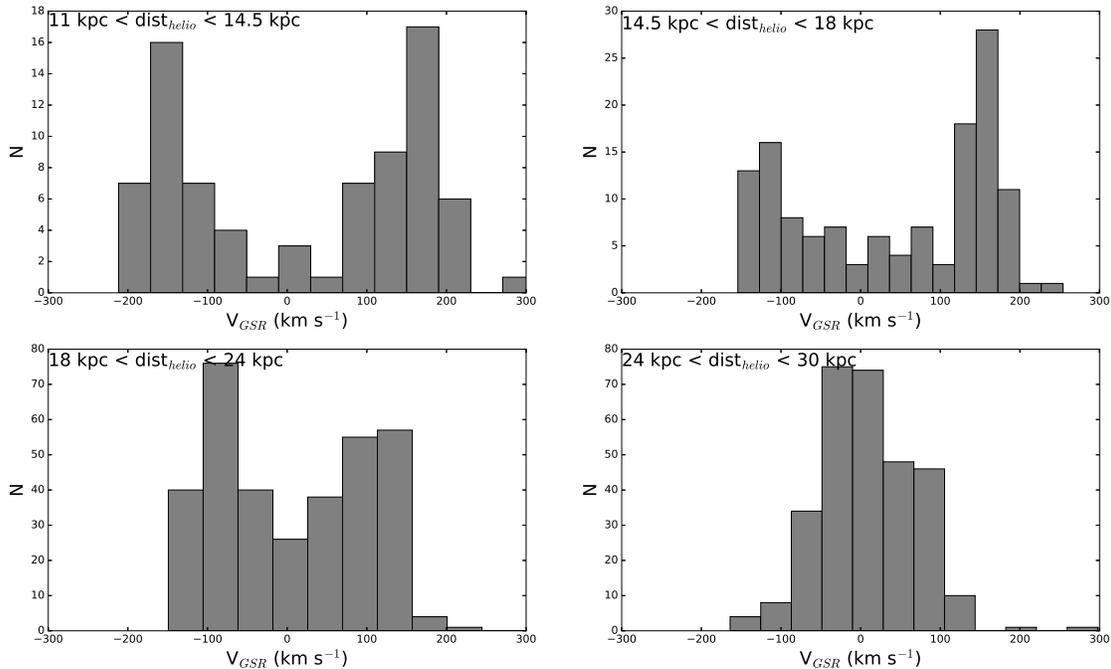}
\caption{ Histogram of line-of-sight, galactic standard-of-rest (V$_{GSR}$) values for a subset of bodies from the Perpendicular Stream N-body simulation in different distance ranges.  Only bodies within the VOD region analyzed in \cite{Newberg2007} (180$^{\circ} <$ RA $< 200^{\circ}$, -10$^{\circ} <$ Dec $<+5^{\circ}$) are shown. Note that at closer distances we see two velocity peaks - one for stars moving away from us towards aphelion (which is close to apogalacticon) and one coming towards us having just passed through aphelion.  As we look at more distant debris in the simulation, histogram appears to be a well-mixed halo population when in fact all of the simulated bodies are part of the same tidal debris substructure.  The panels in this figure can be compared to the histograms of very blue turnoff star ($0.1<(g-r)_0<0.2$) velocities in the bottom two panels of Figure 10 from \cite{Newberg2007}, which show two velocity peaks for very blue turnoff stars with $19.4<g_0<20$ and one broad peak centered at zero for stars with $20<g_0<20.5$. \label{fig:doublehorn}}
\end{figure}

Our orbit for the Perpendicular Stream reaches aphelion near 26 kpc (Figure \ref{fig:nbody}).  This is farther than many previous surveys have $V_{GSR}$ data; for example \cite{Newberg2007} looked in main-sequence turnoff stars from approximately 11 kpc to 18 kpc, observing the double-horned structure at 15 kpc from the Sun.  Our orbit for the Perpendicular Stream predicts the double-valued line-of-sight velocities at these distances, as is demonstrated in Figure \ref{fig:doublehorn}, which shows histograms of the line-of-sight velocities of the Perpendicular Stream N-body data in the Virgo region with a range of heliocentric distances.  Note that histograms of closer stars show the positive and negative velocity peaks (the ``Horns of Virgo").  Closer to the apogalacticon, the distribution looks Gaussian and about the same width as ascribed to a well-mixed stellar halo population.  However, these simulated bodies are from a single merger event.

Since this double-horned structure is often seen in the stars associated with the VOD in previous papers, and the Perpendicular Stream as fit in this paper produces this structure, it is reasonable to conclude that the Perpendicular Stream is responsible for the majority of the material in the VOD.

\subsection{Group D}

Group D stars have similar angular momentum values to those in Groups A \& B, but have quite a bit lower energies.  We fit an orbit to Group D using the same method as was used for Groups A \& B, since it also includes stars with positive and negative line-of-sight velocities as viewed from the Sun.  The orbital parameters are tabulated in Table \ref{tab:orbit-params}.  N-body simulations similar to those described in Section \ref{sec:perpstream} were run on these orbits.

The orbit fit and N-body simulation for Group D can be seen in the second row of Figure \ref{fig:nbody}. This orbit has significant similarities to the orbit fit to Groups A, B, \& H, despite the large difference in energy.  The N-body simulation, generated using the same method as was used in Section \ref{sec:perpstream}, shows the orbit for this moving group going back and forth between the VOD and the HAC regions of the sky, similar to the Perpendicular Stream. The lower energy results in a smaller apogalacticon distance,  but the position of the stream in the sky and the shape of the graph of distance to the orbit as a function of sky position is not unlike that of the Perpendicular Stream.  We will discuss the similarities in more detail in Section \ref{sec:commonorigin}.

\subsection{The Parallel Stream}

Group F also appears similar to Groups A \& B in angular momentum, but has a slightly higher energy.  We use the average values for Group F to determine its orbit, since the member stars are comoving and colocated.  The orbital parameters are completely determined by the 3D position and 3D velocity of the center of this group, and are found in Table \ref{tab:orbit-params}.  An N-body simulation was created on this orbit using the same method as was used in Section \ref{sec:perpstream}.

In the third row of panels from Figure \ref{fig:nbody}, we provide our orbit fit and N-body results for Group F.  Plotted over these results are the orbit fit to the VSS from \cite{Carlin2012}, the selected stars in the Parallel Stream from \cite{Sohn2016}, and the density fit to the Parallel Stream from \cite{Weiss2018b}.  The Group F orbit comes closer to the Sun than the orbit for the VSS (by about 5 kpc), but both orbits are located at the same heliocentric distance in the region of the VOD, where the stars that were used to determine the orbits are located. The orbit fit to Group F lies within the errors given for the orbit from \cite{Carlin2012}. Since these orbits are similar, we suggest that Group F is related to this fit to the VSS, though the stars identified in other papers as VSS stars could be associated with the Perpendicular stream instead.  The orbital parameters for Group F are given in Table \ref{tab:orbit-params}.  We also evolved a dwarf galaxy along the orbit following an identical procedure with an identical Plummer sphere dwarf galaxy and evolution time as for the Perpendicular Stream orbit.  

Note that the higher energy (compared to Groups A \& B) results in a much more distant apogalacticon and an orbit that is more aligned with constant declination in the vicinity of Virgo, rather than constant right ascension like the Parallel Stream.  The orbit is plausibly aligned with the Parallel Stream detections of \citet{Weiss2018b}, and goes right through the location of the \citet{Sohn2016} detection of the Parallel Stream.  The data from \cite{Sohn2016} produces a proper motion measurement of ($\mu_\alpha$, $\mu_\delta$) = (0.48,0.02) mas/yr for the Parallel Stream. The portion of the N-body simulation for Group F, selected within the region 150$^{\circ}$ $<$ RA $<$ 170$^{\circ}$, 0$^{\circ}$ $<$ Dec $<$ 10$^{\circ}$, 30 kpc $<$ dist$_{helio}$ $<$ 35 kpc contains a clump of stars with an average proper motion measurement of ($\mu_\alpha$, $\mu_\delta$) = $\sim$(0.3,0.05) mas/yr.  While the proper motion values from the simulation are not exactly the same as the observed data, they are consistent within the errors of this N-body simulation, as we do not have exact physical constraints for the progenitor's parameters.  This suggests that Carlin's VSS and the Parallel Stream as described in \cite{Sohn2016} are plausibly the same structure.

Around RA = $175^{\circ}$, the N-body data widens to cover a large range of distances and sky area near the VOD, from $\sim 5$ kpc to $\sim 60$ kpc. This variance in distance could be responsible for the discrepancy in our orbit and Carlin's orbit for the VSS, as there is a wide range of material with different velocities which could have been selected. We argue that Group F is also consistent with the Parallel Stream as described in \cite{Weiss2018b} due to the large amount of debris left in the region surrounding that Parallel Stream density fit.  This implies that Carlin's VSS and the Parallel Stream as described in \cite{Weiss2018b} are plausibly the same structure. Note that \cite{Weiss2018b} located the Parallel Stream in a similar region of the sky but across a constant distance, $\sim 15$ kpc, and large width; the measured full width at half maximum is $\sim 13$ kpc.  The reason for the discrepancy in distance trend is unclear, but it might be that: As the N-body simulation of the stream suddently widens substantially right at the position where the Parallel Stream was detected, the algorithm might not have had a good way to determine direction, so instead it fits the mass of stars with a wide, horizontal volume.  \cite{Weiss2018b} may also have found a primarily horizontal stream due to attempting to fit a cylindrical stream model to the VSS while also attempting to fit the Perpendicular Stream in the same region. Since the vertical substructure would be associated with Perpendicular Stream and made unavailable to the Parallel Stream density fit, the algorithm would preferentially select a horizontal orientation for the rest of the cloud.  

The orbits for Group F from this work, the VSS from \cite{Carlin2012} and the fits to the Parallel Stream in \cite{Sohn2016} and \cite{Weiss2018b} all seem to be in similar positions on the sky and heliocentric distances, suggesting a common origin.  The similarity of all of these structures leads us to suspect that they all belong to the same substructure, which we will call the Parallel Stream.

\subsection{Cocytos}

To calculate the orbit for the part of Group H unaffiliated with Groups A \& B, only the four member stars (shown as filled yellow squares in Figure \ref{fig:starplot}) clustered tightly around ($\mu_\alpha$, $\mu_\delta$) = (2, 2.5) mas yr$^{-1}$ were considered. The orbit for this group was simply the average of the velocity and position information for all four stars, as they are colocated and comoving. These stars were also clearly different from the rest of the moving groups based on their $L_x$ and $L_y$ in the bottom right panel of Figure \ref{fig:starplot}. The other 4 member stars in Group H with $L_x$ $>$ -1000 kpc km s$^{-1}$ (shown as filled yellow diamonds in Figure \ref{fig:starplot}) have orbits similar to those the Perpendicular Stream and were discussed in Section \ref{sec:perpstream}.  We evolve an N-body simulation along the orbit of the tight group of four stars following the methodology in Section \ref{sec:perpstream}.  For simplicity, the progenitor properties and evolution time for this simulation is the same as for the other N-body simulations in this paper.

In the bottom two panels of Figure \ref{fig:nbody}, we present the orbit fit and an N-body simulation for Group H.  We show in these panels that the Group H orbit is fairly consistent with the polynomial fit on the sky proposed in \cite{Grillmair2009} for the Cocytos stream.  The polynomial fit to the sky positions of the stream is not to be confused with an orbit fit; this work presents the first ever orbit fit to a Cocytos-like stream derived from kinematic data.  The spread in observed positions of stars in this stream suggest a wider stream than is produced by our N-body.  Note that there is an inconsistency in distance between our orbit and the fit from \cite{Grillmair2009}.  Or orbit shows the distance to Cocytos rapidly changing over the region in which it was detected, while Grillmair's convolution kernel method suggests an orbit with relatively constant distance over the region in which it was detected.  Because we have seen that convolution kernel methods  often produce inaccurate distance estimates \citep[see discussion in][]{Martin2018}, we have chosen to ignore this discrepancy and suggest that the tight group of stars found in Group H could be associated with the Cocytos Stream.

\section{The Virgo Radial Merger: A Common Origin for the Perpendicular and Parallel Streams} \label{sec:commonorigin}



The similarity of the orbit fits to all of the moving groups (excluding the Cocytos Stream), which are all on radial orbits with similar angular momentum but somewhat different energies, leads one to question whether they might all be explained by one merger event.  The orbital parameters in Table \ref{tab:orbit-params} show that the Perpendicular Stream, Parallel Stream, and Group D all pass within 1 kpc of the Galactic center and have eccentricities very close to 1.0, so they are on highly radial orbits.  While the position of the apogalacticon for each of these streams is not identical, the orbits are similar enough that they could plausibly arise from the same progenitor.  If one imagines that a larger dwarf galaxy passes within a scale length of the Galactic center, then a range of energies and angles to apogalacticon could be produced; not all of the dwarf galaxy stars would even necessarily go the same direction around the Galactic center.

\begin{figure}
\center
\includegraphics[width=\linewidth]{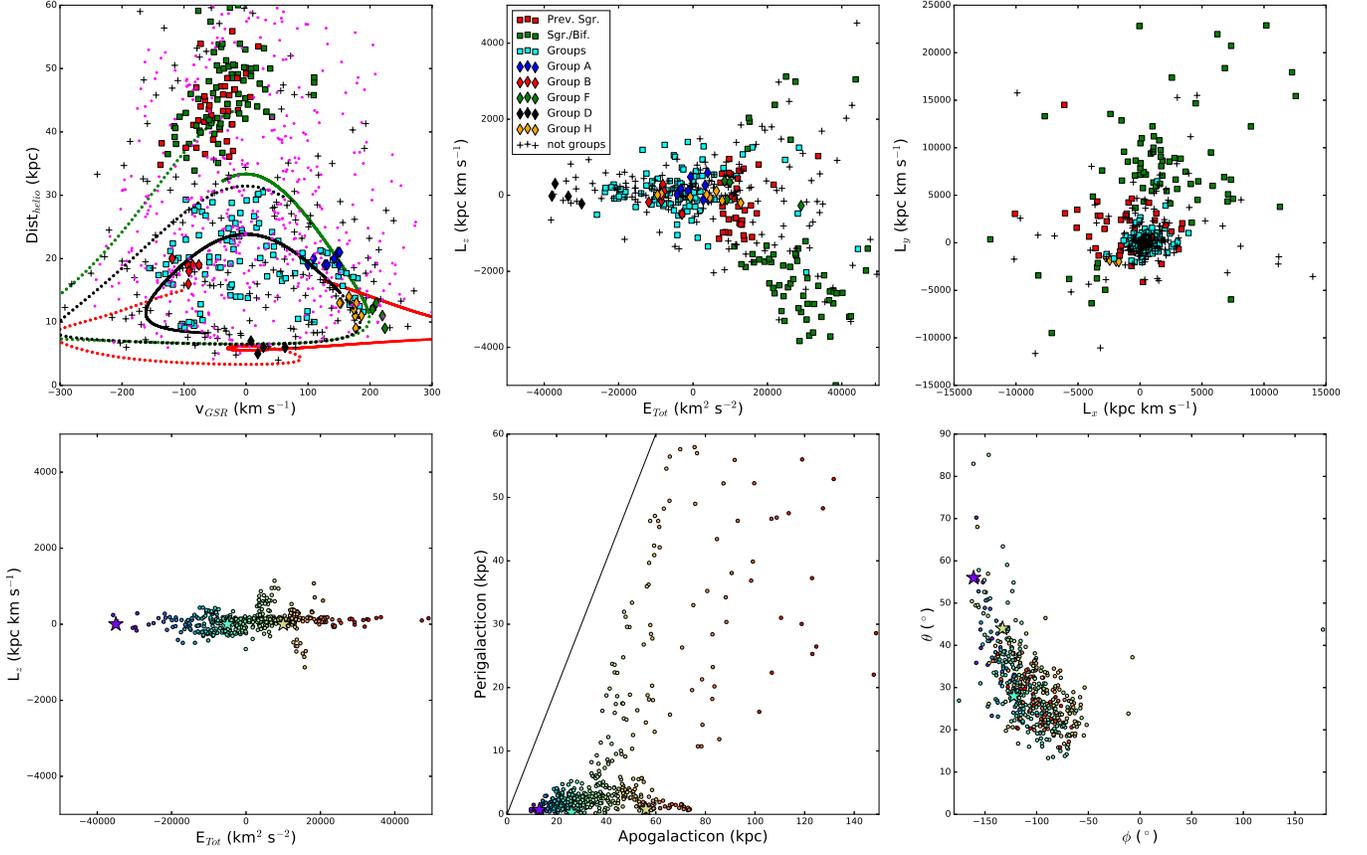}
\caption{ \textit{Top Row:} Kinematic information for the catalog of data in \cite{Vivas2016}. The RRL stars classified as part of the moving groups in \cite{Vivas2016} (squares) are shown in heliocentric distance and $V_{GSR}$. The black crosses are the RRL stars which were not included in any of the moving groups, but were part of the original Vivas catalog. In the top left panel, showing heliocentric distance vs. $V_{GSR}$, we plot the fit orbit for the Perpendicular Stream (black), the orbit for the Parallel Stream (green) and the orbit for Group D (red). The orbits are plotted forward in time for 0.3 Gyr (solid line) and backwards in time for 0.3 Gyr (dashed line). Also plotted are the moving groups from our analysis of \cite{Duffau2014} in diamonds. The majority of moving groups within 30 kpc lie close to the Perpendicular Stream orbit, indicating that it is a large contributor to the density of RRL stars in this region of the sky.  The magenta points show a sample N-body simulation of the Virgo Radial Merger, showing that a single large dwarf galaxy on a radial orbit can produce debris over the entire distance and velocity range in which we see debris in the VOD.  This N-body data is cut to the region of the sky in which the moving groups were detected.  The center panel shows $L_z$ vs. energy for the RR Lyrae and groups from the top left panel.  Note that the most distant groups (red and green squares) separate in angular momentum.  This is also observed in the top right panel, which shows $L_y$ vs. $L_x$.  The green squares have angular momentum expected for the Sgr dwarf tidal stream, while the red squares have angular momentum similar to the cyan squares.  This leads us to believe that the entire set of groups could be the result of one large merger.  Also note that there is a continuous energy spread in the top middle panel for all of the stars with $L_z \sim$ 0, supporting the idea that the individual constituent streams are part of one larger progenitor.\\ 
\textit{Bottom Row:} N-body data from the top left panel colored according to total energy.  Red corresponds to high energy and purple to low energy. The left panel shows angular momentum vs. energy. It shows that the simulation is able to recover a spread in energy similar to that seen in the top middle panel for the data. The lower middle panel presents a comparison of apogalacticon and perigalacticon for the simulated data in the region of the VOD.  As expected, the majority of the orbits are radial (low perigalacticon), and the higher energy orbits have larger apogalacticon. The bottom right panel shows the direction of the apogalaction of the orbits of the simulated debris in coordinates of $\theta$ vs. $\phi$, using the same definitions as is given in the caption for Table \ref{tab:orbit-params}. Also shown (stars) are the $\theta$ vs. $\phi$ values for the Perpendicular Stream (yellow), Parallel Stream (cyan), and Group D (purple), from Table \ref{tab:orbit-params}. These stars are colored according to each structure's total energy. The range of angles of the simulated debris covers the observed angles of the orbits fit to these three substructures, and the energies of each structure are reasonable matches to the energies of the simulated bodies in their respective portions of the diagram. \label{fig:vivas}}
\end{figure}

To explore this, we study the catalog of 412 RRL stars produced by \citet{Vivas2016}, which provides a large collection of stars that extend out to a heliocentric distance of 60 kpc.  \citet{Vivas2016} separated these stars into moving groups in line-of-sight velocity and heliocentric distance using a ``friends-of-friends" algorithm. Many moving groups that appear to actually be a part of the Perpendicular Stream were statistically rejected in the later sections of that work, so we plot all of the groups found in that work before rejection in the top row of Figure \ref{fig:vivas}. The individual moving groups are not distinguished from one another in this figure, as they are shown to all be members of the Perpendicular Stream. The large group beyond a distance of 35 kpc was presumed to be associated with Sgr, though we will later show that this assumption is not completely accurate.

In the top left panel of Figure \ref{fig:vivas}, we plot the heliocentric distance against V$_{GSR}$ for all of the stars in the \cite{Vivas2016} catalog and the stars in our moving groups from \cite{Duffau2014}. Also plotted over this data are the calculated orbits for the Perpendicular Stream (black), the Parallel Stream (green), and Group D (red).  Looking at the stars that \cite{Vivas2016} selected as moving groups in the VOD (filled squares in the figure), we see that the Perpendicular Stream orbit passes through the middle of the cluster of groups. We also see that the Vivas et al. groups fill in the gaps along the orbit between the moving groups found by \cite{Duffau2014}, including groups at the furthest protion of the Parallel Stream, where the line-of-sight, Galactic standard-of-rest velocity is zero. Most of the cyan groups near $V_{GSR}=0$ are at a distance between 20 kpc and 28 kpc, which is consistent with our expectation for the Perpendicular Stream.

The top center panel of Figure \ref{fig:vivas} shows $L_z$ vs. total energy for all of the stars in the \cite{Duffau2014} and \cite{Vivas2016} catalogs.  We used the same Galactic potential model here as we did in calculating the energy for the \citet{Duffau2014} groups.  We see a continuous energy distribution in the cyan RR Lyrae stars from -20,000 to 10,000 km$^2$ s$^{-2}$. The moving groups from \cite{Duffau2014} also extend over this region, excluding Group D.  However, if one looks at the RR Lyrae stars that are not in groups (plus signs), they bridge the gap to Group D.

In this same panel, one sees that the groups that were assumed to be part of the Sgr dwarf tidal stream are spread over a wide range of $L_z$.  Within the RRLs beyond 30 kpc in the top left panel, we have designated two different groups: one red and one green. The red stars are those with $|L_z| < 1500$ kpc km s$^{-1}$, while the green stars are all of the remaining RRLs. The stars represented by red squares appear to be only slightly higher energy than the cyan stars, which we have associated with the Perpendicular Stream. Canonically, Sgr has an $L_z$ of $\sim -3500$ kpc km s$^{-1}$ \citep{Law2005}, indicating that the stars represented by green squares are indeed Sgr and the red stars are another structure. The red stars' integrals of motion suggest that the red squares could be associated with the cyan squares, and the Sgr stream would only explain the green squares.  The top right panel of Figure \ref{fig:vivas} further supports the idea that the groups represented by the green squares have a different origin from that of the red squares, and that the red squares are plausibly associated with the cyan squares; the red squares and blue squares center around $L_x=L_y=0$, while the green squares have primarily positive $L_x, L_y$ angular momenta.

The similarity of the Parallel Stream, Perpendicular Stream, and Group D orbits leads us to question whether they could all be produced by a single radial merger.  This possibility is further suggested by the nearly continuous range of energies of RR Lyrae stars in the top center panel of Figure \ref{fig:vivas}.  To produce the large energy range, we need a much larger dwarf galaxy progenitor than was used to produce the N-body simulations in Figure \ref{fig:nbody}.

The result of an N-body simulation with a progenitor of total mass of 10$^9$ M$_\odot$ and scale radius of 3 kpc, a 2 Gyr evolution time, and 100,000 particles is plotted as magenta dots in the top left panel of Figure \ref{fig:vivas}. The progenitor was integrated along the Perpendicular Stream orbit. Only the bodies that ended up in the part of the sky probed by the RR Lyrae sample $(170^\circ < $RA$ < 200^\circ$, $-5^\circ <$ Dec $< 5^\circ$) were plotted.

These simulated stream particles spread over the entirety of the top left panel of the figure, becoming denser near the proposed location of the Perpendicular Stream. They also extend all the way up to 60 kpc to populate the region even a little more distant than the Sgr dwaf tidal stream.  We are able to replicate the energy range of the radial substructures, using bodies than ended up in just the VOD region of the sky (see lower left panel of Figure \ref{fig:vivas}.  The increasing range of $L_z$ for larger energies found in the data (top center panel) is not reproduced in the model (lower left panel) and is thought to be the result of increasing $L_z$ errors with distance. 

Our success in reproducing the energy range of the Vivas moving groups plotted in cyan \citep[and the moving groups from][]{Duffau2014} suggest that they could all be the result of one large merger. The observed structure could result from a large progenitor that produces debris with a large energy spread due to the range of kinetic energies of the dwarf galaxy stars when it passes near the Galactic center and becomes unbound.  Dynamical friction could also increase the range of energies observed in the tidal debris.

Further evidence that the different streams are consistent with a single merger is shown in the lower center and right panels of Figure \ref{fig:vivas}.  The plot of perigalacticon vs. apogalacticon shows that the simulated data lies on radial orbits with small perigalacticon, and that the apogalacticon increases with energy (as expected).  The lower right plot shows that the position of the apogalacticon of the simulation bodies spans the range of apogalacton positions of the Perpendicular Stream, Parallel Stream, and Group D.  Moreover, the energies of the stars in the groups is consistent with the energies of the bodies with similar apogalacticon positions.

This N-body simulation is only a proof of concept, however; if this simulation were to be plotted on Figure \ref{fig:nbody}, the bodies would significantly deviate from the positions of the orbit due to the size of the progenitor, and would not reproduce the correct sky positions for the moving groups.  The simulation does show that it is possible to populate the range of distances and velocities of the \citet{Vivas2016} RR Lyraes with one dwarf galaxy progenitor.

The N-body simulation does not reproduce the apparent scarcity of RR Lyraes (and groups) starting at about 30 kpc from the Sun.  The increased number of distant stars in groups could be caused in part due to the fact that the friends-of-friends algorithm used in \cite{Vivas2016} preferentially selected groups in regions with a higher density of stars, as is true near the Sgr dwarf tidal stream. However, this does not completely explain the gap. It is difficult to know whether or not the \cite{Vivas2016} catalog suffers from selection effects that may cause this 30 kpc gap.  It is also possible that the gap might be reproduced for an N-body simulation that also reproduces the correct sky positions.  Accurately deriving the orbit for a large dwarf galaxy that passes through the Galactic center and also puts debris on the path of the fitted orbit is a task relegated to a future publication. 

The ratio of BHB stars to RRL stars in the data for the Perpendicular and Parallel Streams also suggests similar compositions for the two structures. 5 out of 15 (33\%) of the stars in the Perpendicular Stream data are BHBs, while 1 out of 4 (25\%) of the stars in the Parallel Stream data are BHBs. This ratio implies a similar stellar population for both substructures, which in turn suggests that these structures may be related. 

Due to the continuous energy spread of the data and the ability to fit the kinematics of the data with a single stream, it appears possible that all of the major substructure (except Sgr and Cocytos) in the VOD region can be attributed to one large merger event. We name this event the Virgo Radial Merger. We show that an evolution time of 2 Gyr is sufficient to create the energy and spatial extent observed in the structure. Note that the N-body simulations put debris in other parts of the sky than the Virgo Overdensity.  Possible associations of the extended debris with known halo substructure is explored in the next section.

\section{Connections to other halo substructure} \label{sec:literature}


\subsection{The Hercules Aquila Cloud \& Eridanus Phoenix Overdensity} \label{sec:hac}

We have seen that the N-body simulations for the Perpendicular Stream, Parallel Stream, and Group D all provide evidence of material being deposited at positions throughout the HAC region. This is consistent with \cite{Simion2019}, which proposed that the VOD and HAC may share a common origin, and is also consistent with the idea that all of these substructures are part of a single Virgo Radial Merger.  Since the stars are on nearly radial orbits, it is expected that stellar debris would preferentially accumulate on both sides of the Galaxy as they reach a apogalacticon on either end of a nearly radial orbit. In this case, the other side of the Galactic Center from the VOD is the HAC. Note that Group D's N-body simulation appears to primarily deposit mass within the HAC region, but also populates the closer portions of the VOD. Interestingly, this group gives us the best match to the canonical distance and extent of the HAC, which is expected to be at 10-25 kpc \citep{Belokurov2007}.

The VOD, the HAC, and the Eridanus-Phoenix Overdensity \citep[EriPhe;][]{Li2016} all lie on a plane that passes through the Galactic center. \citet{Li2016} therefore suggested that EriPhe could be associated with the VOD and the HAC. Interestingly, the N-body simulation for the Perpendicular Stream generates a cluster of material at RA = $\sim$ 60$^\circ$, Dec = $\sim$ -30$^\circ$, which is relatively close (given that the simulated structure is $30^\circ$ across and is very distant from the part of the sky where the data we are trying to recreate is located) to the canonical location of EriPhe at RA = $\sim$ 30$^\circ$, Dec = $\sim$ -55$^\circ$. The material in the N-body data was located at a heliocentric distance extending from 10 kpc to 35 kpc, which extends across the canonical distance to EriPhe of 16 kpc from the Sun.  Note that the position of the 16 kpc material in the simulation is closer to the position found in \citet{Li2016} than the data that is farther from the Sun. Future publications, including the development of a single progenitor N-body that recreates all of the observed data, may provide a more accurate placement of EriPhe in the simulated data.

\subsection{The {\it Gaia}-Enceladus Merger} \label{sec:sausage}

The {\it Gaia}-Enceladus Merger (GEM) \citep{Helmi2018}, also known as the ``Sausage'' Merger \citep{Iorio2019}, is a massive ancient merger that is believed to have been responsible for the majority of the Galactic stellar halo. The GEM is characterized by a wide spread in Galactocentric radial velocity and a flattening in rotational velocity in halo stars within a few kpc of the Sun. This characteristic shape is seen in our data (Figure \ref{fig:starplot}), implying that perhaps our data may be associated with it. 

\begin{figure}
\center
\includegraphics[width=\linewidth]{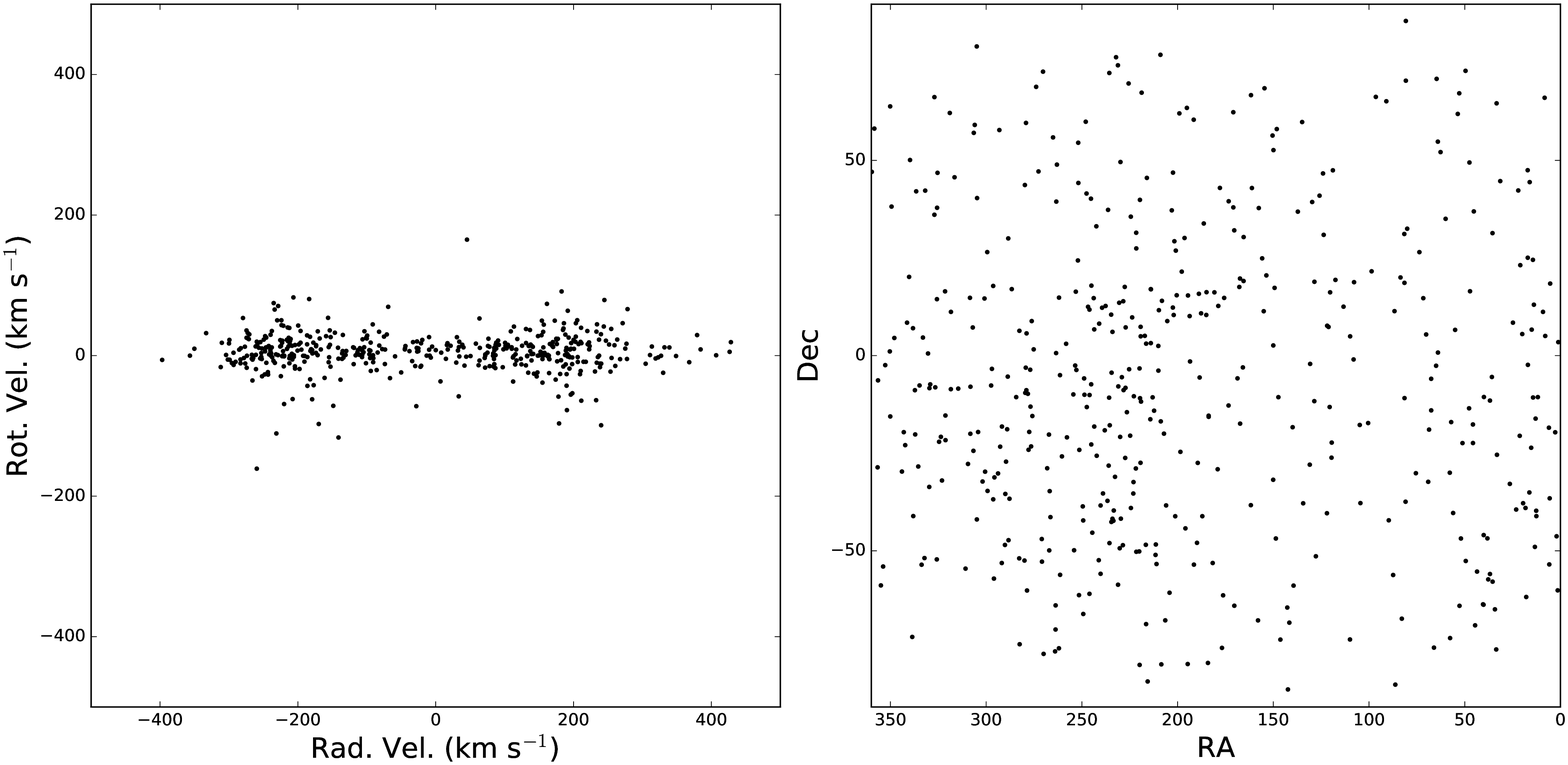}
\caption{\textit{Left:} Galactocentric rotational velocity vs. Galactocentric radial velocity for all stars in the Virgo Radial Merger N-body simulation that are within a distance of 5 kpc of the Sun. These stars show a velocity distribution similar to that of the {\it Gaia}-Enceladus Merger. \\ \textit{Right:} RA vs. Dec for the same stars. This shows that the nearby material from the Virgo Radial Merger would appear as a full sky structure. A nearby full sky structure with this velocity characterization is consistent with the {\it Gaia}-Enceladus Merger. \label{fig:virgogem}}
\end{figure}

To test this, we looked at the bodies in the Virgo Radial Merger simulation that were within 5 kpc of the position of the Sun (Figure \ref{fig:virgogem}).  The simulation put 427 of the 100,000 bodies in this region.  They occupy a similar region of the rotational velocity vs. radial velocity diagram as the stars in the VOD region. They are scattered over the entire sky, as seen from the position of the Sun.  This leads us to believe that the observations of the GEM in the solar neighborhood could be from the same merger as the Virgo Radial Merger, though the relative mass of each structure should be considered in a future work that optimizes the N-body simulation to fit all of the data, including the merger's age.

\cite{Simion2019} claims that the VOD and HAC are remnants of the GEM that are not yet completely phase-mixed into the stellar halo. The GEM is reported to have occured 8 to 11 Gyr ago \citep{Helmi2018}.  We have not tried to rule this merger age out, but the fact that we can make the local debris with a 2 Gyr simulation calls into question the need for a large age.  Long integration times for a large galaxy on the Virgo Radial Merger orbit result in material that is more broadly spread throughout the halo.

\section{Conclusions} \label{sec:conclusions}

In this paper we drew connections between moving groups and tidal streams that had been previously identified in the VOD.  We then showed that the majority of the substructure in VOD could plausibly be explained by a single merger event we call the Virgo Radial Merger.  We then showed that the Virgo Radial Merger could possibly be associated with the HAC, the EriPhe Overdensity and the {\it Gaia}-Enceladus Merger / {\it Gaia} sausage structure.  

We studied moving groups of RRL and BHB stars identified by \citet{Duffau2014}.  Half or more of the stars in each of the five groups (Group A, B, D, F, and H) remained in the moving group, based on their similarity in Cartesian velocity and position, when {\it Gaia} proper motions were included in order to find 3D Cartesian velocities for the stars.  

An orbit fit and N-body simulation of Groups A \& B showed that these groups were on a radial orbit consistent with the Perpendicular Stream.  Group A is moving away from the Sun at the same speed as the majority of stars in the VOD stars.  Group B is moving towards the Sun at similar speeds to the second peak that is usually seen in the line-of-sight velocity histograms of the VOD.  These two groups (containing the most stars) are therefore connected to the Perpendicular Stream and the majority of the matter in the VOD.  Group D was found to lie on a similar but much lower energy radial orbit.  The N-body simulations of all of these groups placed a significant number of stars in the region of the HAC.  The Perpendicular Stream simulation also plausibly placed debris in the general region of the EriPhe Overdensity.

Group F produced a radial orbit that was also similar to the Perpendicular Stream orbit, but with higher energy.  The orbit was similar to the \citet{Carlin2012} orbit for the VSS, within the range of orbits allowed by measurement error.  It also coincided in sky position and proper motion with the Parallel Stream.

Four of the stars in Group H are on an orbit that is reasonably similar to the sky position of the Cocytos Stream found in \citet{Grillmair2009}.  

We present the orbital parameters for each structure in Table \ref{tab:orbit-params}.

Motivated by continuous energy spread over moving groups in \cite{Duffau2014} and \cite{Vivas2016}, we generated a $10^9$ M$_\odot$ N-body simulation of a dwarf galaxy on a radial orbit that can produce stars on orbits similar to the Perpendicular Stream, Parallel Stream, and Group D, including matching their energies, eccentricities, and anglular position of apogalacticon.  This proposed Virgo Radial Merger puts debris out at the distance of the Sgr dwarf tidal stream, and beyond to 60 kpc. In fact, a number of stars previously thought to be associated with Sgr by \cite{Vivas2016} are actually associated with the Virgo Radial Merger, as determined from their angular momentum.  This N-body simulation is only representative, however, and more work needs to be done to create a simulation that simultaneously reproduces all of the observed kinematic and position data.

The Virgo Radial Merger, and in particular the Perpendicular Stream and Group D components, place substantial debris in the region of the HAC.  The Perpendicular Stream also puts debris in a location that can plausibly be associated with the EriPhe Overdensity.  The Virgo Radial Merger simulation places debris in the solar neighorhood that looks reasonably similar to the {\it Gaia}-Enceladus Merger / {\it Gaia} sausage, even though it was only evolved for 2 Gyr.  More work is required to create simulation that matches all of the data and constrains the properties of the progenitor.

\acknowledgments

This work was supported by NSF grant No. AST 16-15688, contributions made by The Marvin Clan, Babette Josephs, Manit Limlamai, the 2015 Crowd Funding Campaign to Support Milky Way Research, and Rensselaer's Summer Undergradaute Research Program.  This work has made use of data from the Sloan Digital Sky Survey. Funding for the Sloan Digital Sky Survey IV has been provided by the Alfred P. Sloan Foundation, the U.S. Department of Energy Office of Science, and the Participating Institutions. SDSS-IV acknowledges support and resources from the Center for High-Performance Computing at the University of Utah. The SDSS web site is www.sdss.org. 
This research also made use of the VizieR catalogue access tool, CDS, Strasbourg, France (DOI : 10.26093/cds/vizier). The original description of the VizieR service was published in A{\&}AS 143, 23. The following VizieR catalogs were used in this work:VizieR Catalogue J/A+A/566/A118 \cite{Duffau2014}, and VizieR Catalogue J/ApJ/831/165 \cite{Vivas2016}.
This work also made use of data from the European Space Agency (ESA) mission {\it Gaia} (\url{https://www.cosmos.esa.int/gaia}), processed by the {\it Gaia} Data Processing and Analysis Consortium (DPAC, \url{https://www.cosmos.esa.int/web/gaia/dpac/consortium}). Funding for the DPAC has been provided by national institutions, in particular the institutions participating in the {\it Gaia} Multilateral Agreement.

\begin{table}[]
\centering

\caption{Orbital parameters for the four structures analyzed in this paper.  Rows of the table give Galactic coordinate position, heliocentric distance, Galactocentric velocities, apogalacticon, perigalacticon, and eccentricity. Also provided are $\theta$ and $\phi$, which are Galactocentric spherical angle coordinates to the position of the next apogalacticon along the orbit, starting from the listed $(l,b,$Dist$_{helio})$ position. $\theta$ is a polar angle and $\phi$ is an azimuthal angle.  The vector pointing from the Sun to the Galactic center is the X-axis; $\theta$ = $\phi$ = 0 along the X-axis. These angles help show the direction of radial motion of each substructure. The parameters are drawn from orbit fits to each significant group outlined in Section \ref{sec:orbits}.\label{tab:orbit-params}}
\begin{tabular}{lcccc}
\hline
\hline
Structure & Perp. Stream (G.'s A \& B) & VSS (Group F) & Group D & Cocytos (Group H)\\

\hline

l ($^\circ$) & 294 & 285 & 289 & 300 \\

b ($^\circ$) & 60 & 61 & 55 & 68 \\

Dist$_{helio}$ (kpc) & 14.6 & 11.3 & 3.1 & 12.5\\

V$_x$ (km s$^{-1}$) & -44 & -135 & -120 & 69\\

V$_y$ (km s$^{-1}$) & -63 & -130 & -54 & 102\\

V$_z$ (km s$^{-1}$) & 145 & 201 & 133 & 215 \\

Apogalacticon (kpc) & 26 & 56 & 13 & 44\\

Perigalacticon (kpc) & 0.3 & 0.8 & 0.7 & 11 \\

Eccentricity & 0.98 & 0.97 & 0.89 & 0.61\\

$\theta ^\circ$ & 28 & 44 & 56 & 54\\

$\phi ^\circ$ & -122 & -133 & -161 & 49\\

\hline
\end{tabular}
\end{table}

\newpage
\bibliographystyle{apj}
\bibliography{references.bib}

\end{document}